\documentclass[a4paper, 11pt]{article}
\pdfoutput=1
\usepackage{jheppubnew}
\usepackage{graphicx}
\usepackage{bm}
\usepackage{ifthen}
\usepackage{amsmath}
\usepackage{amsfonts}
\usepackage{amssymb}
\usepackage{amsthm}
\usepackage{mathrsfs}
\usepackage[all]{xy}
\usepackage{comment}
\usepackage[numbers,sort&compress]{natbib}
 \usepackage[utf8]{inputenc} 

\newcommand{\be}{\begin{equation}}
\newcommand{\ee}{\end{equation}}
\renewcommand{\[}{\begin{equation}}
\renewcommand{\]}{\end{equation}}

\newcommand{\zb}{\bar{z}}

\newcommand{\D}{\mathrm{d}}

\renewcommand{\O}{\mathcal{O}}
 
\newcommand{\lla}{\<\!\<}
\newcommand{\rra}{\>\!\>}

\newcommand{\ep}{\epsilon}

\newcommand{\<}{\langle}
\renewcommand{\>}{\rangle}

\newcommand{\nn}{\nonumber}
\renewcommand{\lla}{\langle \! \langle}
\renewcommand{\rra}{\rangle \! \rangle}

\newcommand{\bs}[1]{\boldsymbol{#1}}
\newcommand{\x}{{\bs{x}}}

\DeclareMathOperator{\Li}{Li}

\title{
Double copy structure and the flat space limit of conformal correlators in even dimensions.}
\author[a]{Arthur E. Lipstein}
\author[b]{and Paul McFadden.}
\affiliation[a]{Department of Mathematical Sciences, Durham University, Durham, DH1 3LE, U.K.}
\affiliation[b]{School of Mathematics, Statistics \& Physics, Newcastle University, Newcastle,  NE1 7RU, U.K.}
\emailAdd{arthur.e.lipstein@durham.ac.uk,
paul.l.mcfadden@ncl.ac.uk}

\begin{document}

\abstract{
We analyse the flat space limit of 3-point correlators in momentum space for general conformal field theories in even spacetime dimensions, and show 
they exhibit a double copy structure similar to that found in odd dimensions. 
In even dimensions, the situation is more complicated because correlators contain branch cuts and divergences which need to be renormalised. 
We describe the analytic continuation of momenta required to extract the flat space limit, and show that the flat space limit is encoded in the leading singularity of a 1-loop triangle integral which serves as a master integral for 3-point correlators in even dimensions. We then give a detailed analysis of the renormalised correlators in four dimensions where the flat space limit of stress tensor correlators are controlled by the coefficients in the trace anomaly.

}
\maketitle


\section{Introduction}

Over the past few decades, the study of scattering amplitudes and conformal correlation functions has revealed remarkable new insight into the structure of quantum field theory and quantum gravity. For example, there is now considerable evidence that scattering amplitudes in quantum gravity can be computed from the  correlation functions of a quantum field theory in one lower  dimension. This holographic correspondence is best understood when the bulk geometry is anti-de Sitter \cite{Maldacena:1997re}, and conformal field theory (CFT) correlators in the boundary can be computed from Witten diagrams in the bulk.  In the flat space limit, these correlators reduce to scattering amplitudes in one higher dimension 
\cite{Penedones:2010ue}.  
Similar methods can also be applied to compute cosmological observables \cite{Maldacena:2011nz,Bzowski:2012ih,Arkani-Hamed:2018kmz}.

Since scattering amplitudes arise from the flat space limit of correlators, they are far simpler objects and many more tools are available to compute them. It is therefore of great interest to understand how to generalise these tools to correlators, and there has  been important progress in this direction. For example, techniques analogous to BCFW recursion \cite{Britto:2005fq} and unitarity methods \cite{Bern:1994zx,Bern:1994cg} for scattering amplitudes have  been proposed for correlators \cite{Raju:2010by, Raju:2011mp, Raju:2012zr, Alday:2017vkk, Rastelli:2017udc, Zhou:2018sfz}. Another remarkable property of scattering amplitudes is a set of relations connecting gauge to gravitational amplitudes known collectively as the double copy (see \cite{Bern:2019prr} for a recent review).  Recently,  analogous double copy relations were found for general conformal correlators in odd spacetime dimensions \cite{Farrow:2018yni}. In particular, Euclidean 3-point correlators of stress tensors, conserved currents and marginal scalars were shown to reduce to gauge and gravitational scattering amplitudes in one higher dimension in the flat space limit. This was achieved by working in 
 momentum space and taking the energy (defined as the sum of the magnitudes of the three momenta) to zero.  In three dimensions, certain aspects of this double copy structure even extend
 beyond the flat space limit.

In odd dimensions, 3-point CFT correlators are rational functions of the momentum magnitudes which exhibit poles in the energy.  The scattering amplitudes can then be read off from the coefficients of the most singular poles.  In even dimensions, the situation is more subtle because the correlators contain branch cuts and need to be analytically continued before taking the flat space limit. Our strategy will be to analyse first the flat space limit of a certain 1-loop triangle integral.  All the correlators we consider can then be constructed by applying differential operators to this master integral.\footnote{These correlators are non-perturbative,  being fixed by conformal symmetry.}
 If the energy of each particle is taken to be positive, 
the master integral 
is non-singular as the total energy tends to zero.  
To reach the flat space limit, we must instead analytically continue at least one of the energies to be negative before sending their sum to zero.  
This continuation involves crossing certain branch cuts giving rise to a new term with the desired singular behaviour in the flat space limit. Interestingly, this new term is precisely the leading singularity of the 1-loop triangle integral computed  decades ago by Cutkosky \cite{Cutkosky:1960sp}. We present a more modern derivation of this result by first mapping the triangle integral to a box integral with a remarkable property known as dual conformal invariance, and then evaluating the leading singularity of this box integral by taking the global residue.   

The flat space limit of correlators in general even dimensions can then be deduced by applying the appropriate differential operators to the master integral, and we discover the same double copy structure that we previously found in odd dimensions. In even dimensions there is one further complication coming from the fact that correlators are divergent and need to be renormalised. This renormalisation has been worked out explicitly in four dimensions \cite{Bzowski:2017poo,Bzowski:2018fql}, and we carefully verify our general arguments in this case. We also find that the coefficients of the scattering amplitudes which arise in the flat space limit of stress tensor correlators are controlled by conformal anomalies, in agreement with general holographic expectations \cite{Henningson:1998gx,Nojiri:1999mh,Bugini:2016nvn}. 

The structure of this paper is as follows. In section \ref{review}, we review some basic results about scattering amplitudes and conformal correlators in momentum space that will be relevant for this paper. In section \ref{master} we derive the flat space limit of the master integral after analytic continuation, and in section \ref{general} we use this result to deduce the flat space limit of correlators of stress tensors, currents and marginal scalars in general even dimension by applying certain differential operators.  This reveals double copy structure similar to that previously found in odd dimensions. In section \ref{4d}, we specialise the discussion to four dimensions where the renormalised correlators have been explicitly computed and we verify the general arguments of the previous section. We also show how the anomaly coefficients parametrise the flat space limit. 
We present our conclusions and future directions in section \ref{concl}. In Appendix \ref{Leadingsing} we compute the leading singularity of the master integral.

\section{Review}  \label{review}

In this section, we  review some results about momentum-space conformal correlators in $d$ Euclidean dimensions \cite{Bzowski:2013sza, Bzowski:2017poo, Bzowski:2018fql}, and their relation to scattering amplitudes in $(d+1)$-dimensional Minkowski space, which for odd $d$ were worked out  in \cite{Farrow:2018yni}. The tensor structure of correlators is first decomposed into a basis of  transverse traceless tensors, where each component is multiplied by a scalar form factor.  
 For 3-point correlators, these form factors are functions purely of the momentum magnitudes, 
 \[
 p_i=+\sqrt{\bs{p}_i^2}, \qquad i \in \left\{ 1,2,3\right\},
 \]
 since momentum conservation 
 allows us to replace $\bs{p}_1\cdot\bs{p}_2 = (p_3^2-p_1^2-p_2^2)/2$, {\it etc.}
For physical kinematics, these magnitudes also obey the triangle inequalities $0 \leq p_i \leq p_j+p_k$. 

If desired, the non-transverse traceless parts of correlators can be recovered from lower-point functions via the trace and transverse Ward identities.
Here, since our interest is in scattering amplitudes, we will instead contract all indices with transverse polarisation vectors $\bs{\ep}_i = \bs{\ep}(\bs{p}_i)$ satisfying
\[
\bs{\ep}_i\cdot \bs{p}_i = 0, \qquad
\bs{\ep}_i\cdot\bs{\ep}_i = 0.
\]
Inserting this tensorial decomposition into the conformal Ward identities, one finds the form factors are  given by specific  linear combinations of triple-$K$ integrals \cite{Bzowski:2013sza}, 
\begin{align}
I_{\alpha \{ \beta_1, \beta_2, \beta_3 \}}(p_1, p_2, p_3) & = \int_0^\infty \mathrm{d} x \: x^\alpha \prod_{i=1}^3 p_i^{\beta_i} K_{\beta_i}(p_i x), \label{tripleK} 
\end{align}
where $K_{\beta_i}$ is a modified Bessel function of the second kind. 

To connect with scattering amplitudes, we first lift to $(d+1)$-dimensional Minkowski space by introducing the bulk null momenta and polarisation vectors
\[
p^\mu_i = (p_i,\bs{p}_i), \qquad \ep_i^\mu = (0,\bs{\ep}_i).
\]
For gravitons, we write polarisation tensors in terms of polarisation vectors as $\epsilon^{\mu \nu}_i=\epsilon^\mu_i \epsilon^\nu_i$. 
Contractions of polarisation vectors can then be lifted to their bulk counterparts by replacing $\bs{\ep}_i\cdot \bs{p}_j \rightarrow \ep_i\cdot p_j$ and $\bs{\ep}_i\cdot\bs{\ep}_j\rightarrow \ep_i\cdot\ep_j$.  However, while $d$-dimensional momentum is conserved, the bulk momentum is not since
\[
\sum_{i=1}^3 p_i^\mu = (E, \bs{0}),
\]
where the total bulk energy 
\[\label{Edef}
E=p_1+p_2+p_3.
\]  
We are therefore interested in extracting the leading behaviour of CFT correlators in the limit $E\rightarrow 0$ for which energy conservation is restored. 

This limit  is naturally regarded as a {\it flat space} limit, either of $(d+1)$-dimensional anti-de Sitter space \cite{Penedones:2010ue, Gary:2009ae, Raju:2012zr}, or alternatively of $(d+1)$-dimensional de Sitter space \cite{Maldacena:2011nz, Arkani-Hamed:2015bza, Farrow:2018yni, Arkani-Hamed:2018kmz}.\footnote{The  relation between 3-point correlators in AdS and dS is also known from  holographic cosmology \cite{Maldacena:2002vr, McFadden:2010vh,McFadden:2011kk}, though we will not use this here.}  
This follows since
the leading behaviour as $E\rightarrow 0$ is governed by the asymptotic behaviour of modes deep in the interior of AdS, or equivalently at very early times in de Sitter space, where the effects of spacetime curvature can be neglected.

For CFTs in odd spacetime dimensions, the triple-$K$ integrals feature half-integer indices and the form factors are simple rational functions of the momentum magnitudes.   Taking the flat space limit is then simply a matter of extracting the leading behaviour as $E \rightarrow 0$.  The coefficients of the leading divergences are  $(d+1)$-dimensional flat space scattering amplitudes which exhibit  double copy structure. In  \cite{Farrow:2018yni}, we found that 3-point correlators of stress tensors and currents reduce to linear combinations of the following gauge and gravitational amplitudes,  which are related to each other by a double copy: 
\begin{equation}
\mathcal{A}_{EG}=(\mathcal{A}_{YM})^{2},\qquad \mathcal{A}_{\phi R^{2}}^{222}=\mathcal{A}_{F^{3}}\mathcal{A}_{YM},\qquad \mathcal{A}_{W^{3}}=(\mathcal{A}_{F^{3}})^{2}.
\label{gravityamp}
\end{equation}
Here, $\mathcal{A}_{EG}$ is the 3-graviton scattering amplitude for Einstein gravity, $ \mathcal{A}_{W^{3}}$ is that for Weyl-cubed gravity, while $ \mathcal{A}_{\phi R^{2}}^{222}$ is the 3-graviton amplitude (indicated by the $222$ superscript) for the curvature-squared theory of gravity coupled to scalars constructed in \cite{Broedel:2012rc}.\footnote{In $d=4$, this theory reduces to a certain non-minimally coupled version of conformal gravity \cite{Johansson:2017srf}.}
As indicated, these gravitational amplitudes are double copies of the gauge theory amplitudes
\begin{equation}
\mathcal{A}_{YM}=\epsilon_{1}\cdot\epsilon_{2}\,\,\epsilon_{3}\cdot p_{1}+\mathrm{cyclic},\qquad \mathcal{A}_{F^{3}}=\epsilon_{1}\cdot p_{2}\,\,\epsilon_{2}\cdot p_{3}\,\,\epsilon_{3}\cdot p_{1},
\label{gaugeamp}
\end{equation}
where $\mathcal{A}_{YM}$ is the 3-gluon Yang-Mills amplitude and  $\mathcal{A}_{F^{3}}$ is the corresponding amplitude in a higher-derivative gauge theory with an $F^3$ interaction constructed in \cite{Johansson:2017srf}. It is also natural to consider 3-point CFT correlators involving marginal scalars.  In \cite{Farrow:2018yni}, we found the correlator of two stress tensors and a marginal scalar reduces in the flat space limit to the amplitude
\[
\mathcal{A}_{\phi R^{2}}^{220}=\left(\mathcal{A}_{\phi F^{2}}\right)^{2},
\label{ymdc}
\]
where
\begin{equation}
\mathcal{A}_{\phi F^{2}}=\epsilon_{1}\cdot p_{2}\,\epsilon_{2}\cdot p_{1}.
\label{dilaton}
\end{equation}
Here, $\mathcal{A}_{\phi R^{2}}^{220}$ is the scattering amplitude of two gravitons  and a scalar (indicated by the superscript $220$) in the $\phi R^2$ theory, which is a double copy of the amplitude $\mathcal{A}_{\phi F^{2}}$ for two gluons and a scalar in a Yang-Mills dilaton theory.

In even spacetime dimensions, a more detailed analysis is required in order to extract the flat space limit.  
The two issues are that, firstly, the form factors for CFTs in even dimensions diverge introducing the additional complication of regularisation and renormalisation; and secondly, the resulting renormalised form factors have a more complicated analytic structure containing branch cuts. 
As a result, the  nature of the analytic continuation required to take the flat space limit $E\rightarrow 0$ must be carefully specified.  This is the central question we address in this paper.

\section{Flat space limit of the master integral} \label{master}

\begin{figure}
\centering
	       \includegraphics[scale=1.0]{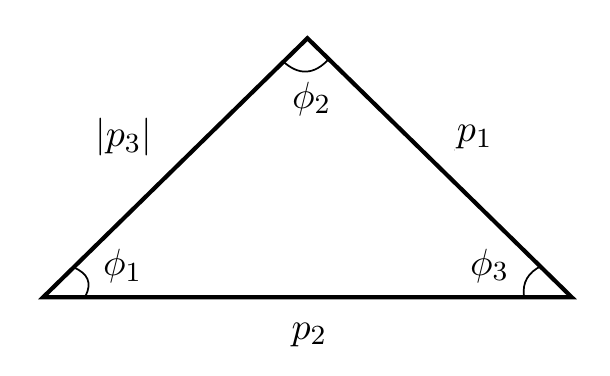}
    \caption{The momenta in the 3-point function form a triangle by momentum conservation, with angle $\phi_i$ appearing opposite the side of length $|p_i|$.} 
    \label{triangle}
\end{figure}

As we will review later in section \ref{reltoI1000}, for even-dimensional correlators all form factors can be obtained recursively starting from the  triple-$K$ integral $I_{1\{000\}}$.  
Our first task, therefore, is to  evaluate the flat space limit of this master integral.  We will discuss this from several points of view, but our basic 
 strategy will  be to analytically continue the momentum magnitude
\begin{equation}
p_{3} =  |p_{3}| \,e^{i\theta}, \qquad 0\le \theta\le \pi,
\label{continuation}
\end{equation}
where the momenta are ordered so that $p_3$ is the largest  magnitude.
After continuing  from $\theta=0$ to $\theta=\pi$, 
the flat space limit $E = p_1+p_2+p_3 \rightarrow 0$ 
then corresponds to sending
\[
%
|p_{3}|\rightarrow p_{1}+p_{2}.
\]
Noting that 
\begin{equation}
K_{\nu}(e^{i\pi}x)=e^{-i\pi\nu}K_{\nu}(x)-i\pi I_{\nu}(x),\qquad x\in\mathbb{R}^{+}, \qquad \nu\in\mathbb{Z},
\label{Kcontinuation}
\end{equation}
where $I_{\nu}$ is a modified Bessel function of the first kind,  we immediately obtain the following expression for the analytic continuation of $I_{1\{000\}}$:
\[
I_{1\{000\}}(p_{1},p_{2},p_{3})= I_{1\{000\}}(p_{1},p_{2},|p_{3}|)-i\pi\int_{0}^{\infty}\D x\, x\, K_{0}(p_{1}x)K_{0}(p_{2}x)I_{0}(|p_{3}|x).
\]
The first term on the right-hand side is simply the original triple-$K$  integral and  is 
 finite in the flat space limit as we will see shortly. The second term can be evaluated using the formula \cite{Gervois:1985ff} 
\begin{align}
&\int_{0}^{\infty}\D x\,x^{1+\mu}K_{\nu}(p_{1}x)K_{\nu}(p_2 x)I_{\mu}(|p_3| x)
\nn\\&\quad 
=
2^{-2\mu-2}\sqrt{\frac{\pi}{2}}\Gamma(1+\mu+\nu)\Gamma(1+\mu-\nu)\frac{|c_{123}|^{\mu}}{\Delta^{2 \mu+1}}\left(\sin\phi_{3}\right)^{\mu+1/2}P_{\nu-1/2}^{-\mu-1/2}\left(\cos\phi_{3}\right),
\label{IKKintegral1}
\end{align}
where 
\[
c_{123}=p_1 p_2 p_3,
\] 
the $P^\mu_\nu$ are Legendre functions and $\Delta$ is the area of the triangle spanned by the momenta as depicted in Figure \ref{triangle}. 
Moreover, using Heron's formula, the area can be written as
\begin{equation}
\Delta=\frac{1}{2}|p_{i}||p_{j}|\sin\phi_{k}=\sqrt{J^{2}}/4,
\label{delta}
\end{equation}
where $\phi_i$ are the angles of the triangle in Figure \ref{triangle} and
\[\label{Jsqdef}
J^{2}=E\,
\left(p_{1}+p_{2}-p_{3}\right)\left(p_{1}-p_{2}+p_{3}\right)\left(-p_{1}+p_{2}+p_{3}\right).
\]
Notice the value of $J^2$ is the same at the start and end-point of our analytic continuation.
The formula \eqref{IKKintegral1} is valid for $1+\mu-|\nu|>0$,
so choosing $\mu=\nu=0$  we find
\[
I_{1\{000\}}(p_{1},p_{2},p_{3})= I_{1\{000\}}(p_{1},p_{2},|p_{3}|)-\frac{i \pi \phi_{3}}{4\Delta}.
\]
Hence, after analytic continuation to $\theta=\pi$, the master integral $I_{1\{000\}}$ acquires a new term which is simply the ratio of the angle $\phi_3$ (opposite to the side $p_3$) to the area of the triangle. In the flat space limit, the angle $\phi_{3}\rightarrow\pi$ and the area of the triangle vanishes according to 
\[
4\Delta=\sqrt{J^{2}}\rightarrow \sqrt{8|c_{123}|E} 
\]
Hence, we find that
\begin{equation}
\lim_{E \rightarrow 0} I_{1\{000\}}(p_{1},p_{2},p_{3})\rightarrow-\frac{i\pi^{2}}{\sqrt{8|c_{123}|E}} = \frac{\pi^2}{\sqrt{-J^2}},
\label{flat1000}
\end{equation}
where the positive sign is taken in the square roots.

Further insight can be obtained by re-deriving this result from a different point of view. 
First, we map the momenta to the complex plane according to 
\[
u=\frac{p_1^2}{p_3^2} = z\bar{z}, \qquad
v=\frac{p_2^2}{p_3^2}= (1-z)(1-\bar{z}).
\]
Choosing $\Im(z)\ge 0$, we can invert to find
\begin{align}
z&=\frac{1}{2}\left(1+u-v+\sqrt{(1+u-v)^{2}-4u}\right),\\[0ex]
\bar{z}&=\frac{1}{2}\left(1+u-v-\sqrt{(1+u-v)^{2}-4u}\right).
\end{align}
For physical momentum configurations satisfying the triangle inequalities $p_i+p_j\ge p_k$, the quantity under the square root is negative ({\it i.e.,} $-J^2/p_3^{4}\le 0$)  meaning $z$ and $\zb$ are complex conjugates.
For such momenta, 
the master integral $I_{1\{000\}}$ is equivalent to a 1-loop triangle integral \cite{Bzowski:2013sza, Bzowski:2015yxv}
\begin{equation}
I_{1\{000\}}(p_{1},p_{2},p_{3})=\frac{1}{4\pi^{2}}\,\int\frac{\D^{4}{\ell}}{\ell^{2}({\ell}+{p}_{1})^{2}({\ell}-{p}_{3})^{2}},
\label{triangleint}
\end{equation}
which can be evaluated in terms of $z$ and $\zb$ as 
\begin{align}\label{I1000zform}
 I_{1\{000\}}& 
 = \frac{1}{2p_3^2(z-\bar{z})}\Big[\mathrm{Li}_2\, z - \mathrm{Li}_{2} \, \bar{z}
+\frac{1}{2}\ln\, (z\bar{z}) \ln\Big(\frac{1-z}{1-\bar{z}}\Big)\Big],
\end{align}
where $\Li_2$ is the dilogarithm.  In fact,
$ I_{1\{000\}}$ is simply the Bloch-Wigner function  \cite{Zagier:2007knq} divided by  
\[\label{Jsqz}
\sqrt{-J^2} =p_3^2 (z-\zb).
\] 
Geometrically, the Bloch-Wigner function expresses the volume of an ideal tetrahedron ({\it i.e.,} with vertices at $0$, $1$, $z$ and $\infty$) living  in the hyperbolic 3-space spanned by the complex $z$-plane times the real line.  Since $\sqrt{-J^2}$ is 
proportional to the area of the Euclidean triangle in Figure \ref{triangle},  the master integral $I_{1\{000\}}$ is thus given by the ratio of these quantities.

 The Bloch-Wigner function has the special property that all  branch cuts in the logarithms and dilogarithms cancel, rendering $ I_{1\{000\}}$ single-valued everywhere in the complex plane. 
Taking the limit $z\rightarrow \zb$ then corresponds to taking the collinear limit $p_1+p_2-p_3\rightarrow 0$, where $J^2$ vanishes according to \eqref{Jsqdef}.\footnote{Recall we ordered our momenta so $p_1+p_2\ge p_3$, hence this specific collinear limit is selected.}
In this collinear limit, the master integral is finite since the pole in $z-\zb$ is cancelled by the vanishing of the numerator \cite{Chavez:2012kn}.

\begin{figure}
\centering
   (a)  \includegraphics[scale=.31]{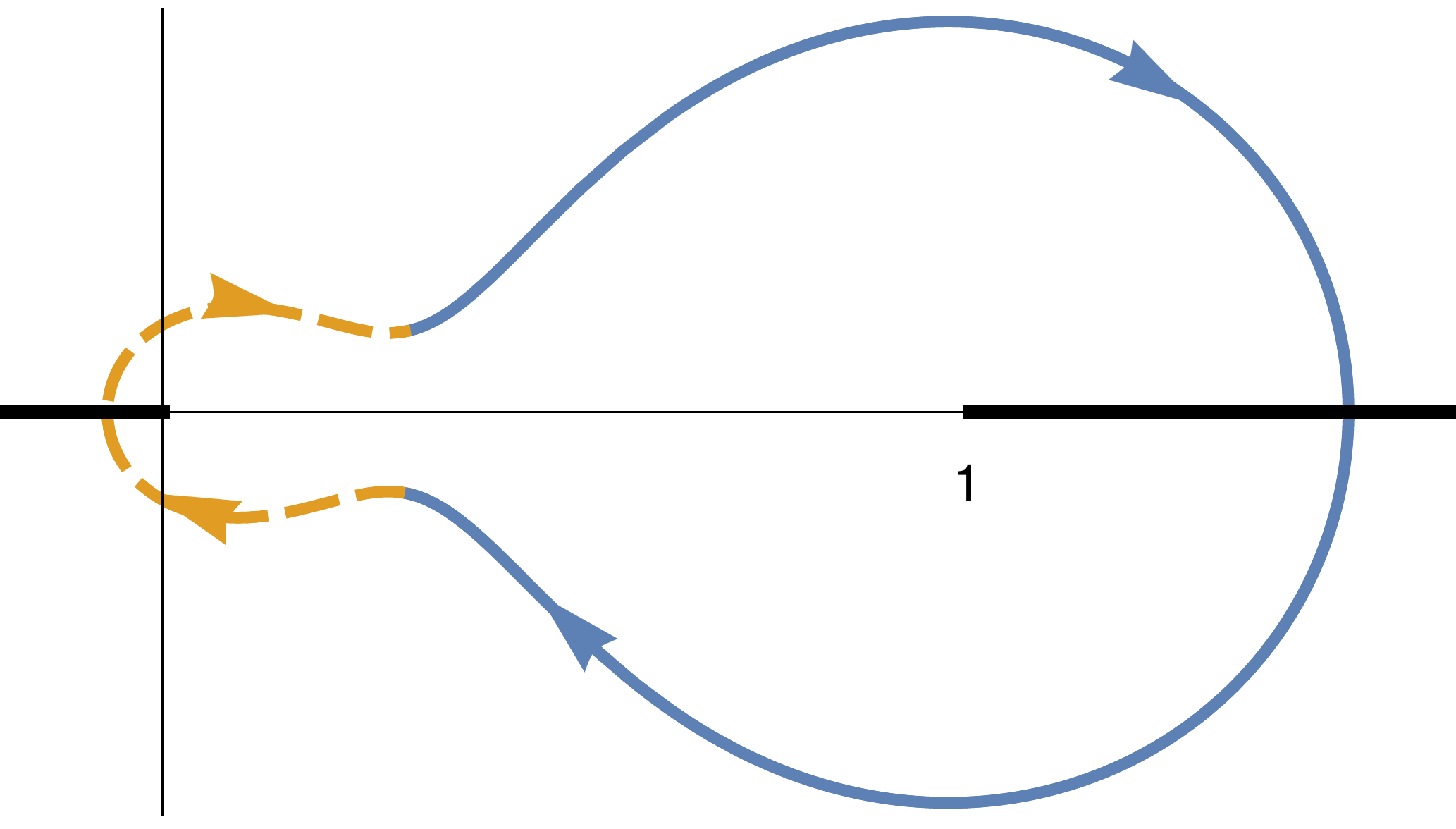}\hspace{0.5cm}
	 (b)      \includegraphics[scale=.31]{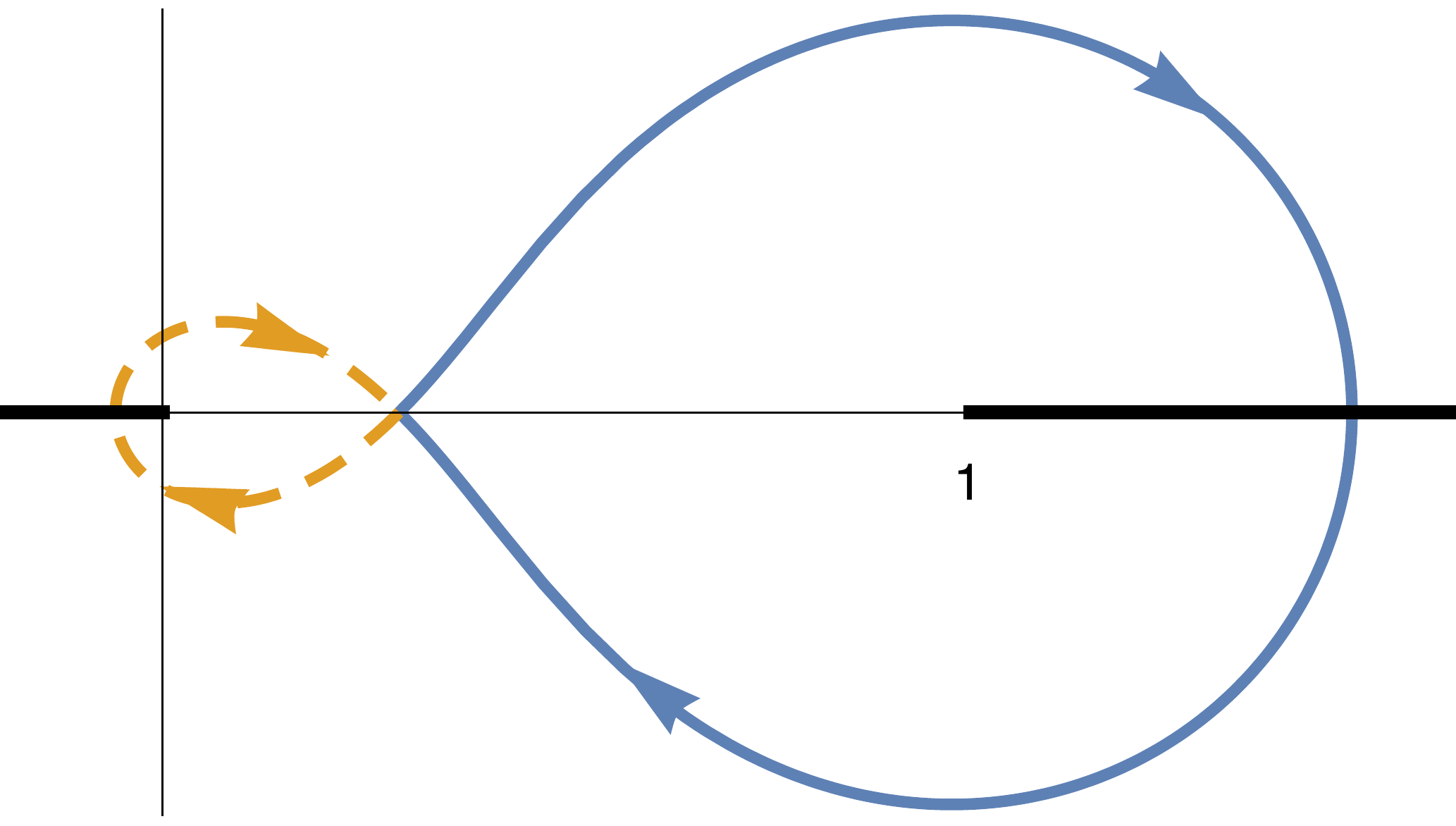}
	     
    \caption{(a) As we increase $\theta$ from $0$ to $\pi$, $z$ and $\bar{z}$ move clockwise in the complex plane following  the solid blue and  orange dashed paths respectively.  Starting from generic complex conjugate initial values (corresponding to physical momentum configurations), they ultimately end up exchanging positions. In the process, $z$ crosses the branch cut between  $(1,\infty)$ while $\bar{z}$ crosses the branch cut between $(-\infty,0)$.  (b) The trajectory of $z$ and $\bar{z}$ as we continue from an collinear initial configuration  to  one with $E=0$. The flat space limit thus corresponds to bringing $z$ and $\bar{z}$ to a point on the real axis between $0$ and $1$ after crossing the cuts in the direction shown.} 
    \label{contour}
\end{figure}

To take instead the {\it flat space} limit $E\rightarrow 0$, we have to first continue $z$ and $\zb$ such that these variables are no longer complex conjugate to  one another.  
The way to do this follows from the continuation of $p_3$ in  \eqref{continuation}, which sends
\[
u = |u| e^{-2i\theta}, \qquad v = |v|e^{-2i\theta}.
\]
As $\theta$ ranges from zero to $\pi$, 
the trajectories of $z$ and $\zb$ are then as plotted in Figure \ref{contour}.   
Starting from complex conjugate initial values, for $0<\theta<\pi$, one finds 
$z$ and $\zb$ are no longer complex conjugates meaning the branch cuts in the logarithms and dilogarithms no longer cancel.   From \eqref{I1000zform}, these cuts are located along the negative real axis, and along the positive real axis for values greater than unity.\footnote{Other placements of these cuts are possible, but the final result is the same.}
 As we increase $\theta$, we find $z$ crosses the branch cut on the positive real axis, while $\zb$ crosses the branch cut on the negative real axis, both in a clockwise sense. Upon reaching $\theta=\pi$ their values are once again complex conjugates, but their final positions are now exchanged relative to their initial ones.    While the exact shape of the trajectory depends on the initial values, the manner in which the respective cuts are crossed is always the same.

As a result of crossing these cuts, we acquire the following new contributions, whose signs are fixed by the direction in which the cuts are crossed:
\begin{align}
&\Li_2(z) \rightarrow \Li_2(z) - 2\pi i \ln z,\qquad 
\ln (1-z) \rightarrow \ln (1-z) + 2i\pi, \qquad
\ln \zb \rightarrow \ln \zb+2i\pi.
\end{align}
After analytic continuation, we therefore find
\[
I_{1\{000\}}\Big|_{\theta=\pi} = I_{1\{000\}}\Big|_{\theta=0} +\frac{1}{2p_3^2(z-\zb)}\Big[i\pi\Big( \ln \Big(\frac{\zb}{z}\Big) +\ln\Big(\frac{1-z}{1-\zb}\Big)\Big) - 2\pi^2\Big],
\]
where the $z$ and $\zb$ on the right-hand side refer to their final values at $\theta=\pi$.\footnote{For the first term this distinction is immaterial since $I_{1\{000\}}\big|_{\theta=0}$ is even under exchanging $z$ and $\zb$.} 
If we now send $z\rightarrow \zb$ so that they collide on the real axis between zero and unity (without crossing any further cuts), we obtain the leading singular behaviour
\[\label{zflatlimit}
\lim_{E\rightarrow0}I_{1\{000\}}\Big|_{\theta=\pi}=-\lim_{z\rightarrow\bar{z}}\frac{\pi^{2}}{p_{3}^{2}(z-\zb)}=+\frac{\pi^{2}}{\sqrt{-J^{2}}}=-\frac{i\pi^{2}}{\sqrt{8|c_{123}|E}}.
\]
The  positive sign in the penultimate equation follows because $z$ and $\zb$ have exchanged positions relative to their initial values in \eqref{Jsqz}.  The value of $\sqrt{-J^2}$ is the same at $\theta=0$ and $\theta=\pi$,  and was fixed as the positive root
 $\sqrt{-J^2} = +i\sqrt{8|c_{123}|E}$ 
by our choice $\Im (z)\ge 0$ for $\theta=0$.

We thus recover the same flat space limit as in \eqref{flat1000} above, and the answer will clearly be the same for any analytic continuation so long as the cuts are traversed in the same manner  before sending $z\rightarrow \zb$.    Had we continued in the opposite sense ({\it i.e.,} with $\theta$ running from zero to $-\pi$) the flat space limit would take the opposite sign, but our real interest here is only in the momentum dependence. 
Continuing from $\theta=0$ to $\pi$ nevertheless seems more natural since this preserves $\Im(z)\ge 0$ for the first part of the trajectory in cases where $z$ and $\zb$ are initially collinear, as illustrated in the right-hand panel of Figure \ref{contour}.

Remarkably, the analytic continuation of this same integral has been studied in the context of  position-space 4-point functions in \cite{Gary:2009ae,Maldacena:2015iua}, where the general behaviour is very similar to that found here. 
The basis for this connection is explained in Appendix \ref{leadingsing_app}, where we relate the  
3-point master integral $I_{1\{000\}}$ to a dual conformal 
box integral. Taking the flat space limit of $I_{1\{000\}}$ then corresponds to computing the leading singularity of this box integral via its global residue. In fact, this connection between the flat space limit and the leading singularity is also visible at the level of the 3-point function.  In \cite{Cutkosky:1960sp},  
Cutkosky showed that the leading singularity of the triangle integral  \eqref{I1000zform}, obtained by putting all three propagators on shell,  has the beautiful geometrical interpretation 
\begin{equation}
\int \D^{4}\ell\,\delta(\ell^{2})\,\delta\left((\ell+p_{1})^{2}\right)\delta\left((\ell-p_{3})^{2}\right)=\frac{\pi}{8\Delta},
\label{leadsing}
\end{equation}
where $\Delta$ is the area of the triangle \eqref{delta}.
Including the factor of $(2\pi i)^3$ accompanying the delta functions, and the factor of $1/4\pi^2$ in \eqref{triangleint}, we again recover precisely \eqref{flat1000}.

\section{
Evaluating the flat space limit of correlators}
 \label{general}

Having  
analysed the master integral, let us now 
discuss how to evaluate the flat space limit of correlators for general even  dimensions $d=2n\ge 4$. 
The relevant correlators for the double copy are $\<JJJ\>$ and $\<TTT\>$, as well as $\<JJ\O\>$ and $\<TT\O\>$ for a marginal operator $\O$.  
As for the master integral, a number of different approaches can be taken.
We discuss these in each of the following three subsections.  
While all approaches give the same result, they 
 present different features of interest.

\subsection{Asymptotic analysis}

First we present a simple asymptotic formula for the flat space limit of a general triple-$K$ integral.  
We start by analytically continuing the general triple-$K$ integral \eqref{tripleK}.   Using \eqref{continuation}  and \eqref{Kcontinuation}, we obtain
\begin{align}
I_{\alpha\{\beta_{1}\beta_{2}\beta_{3}\}}(p_{1},p_{2},p_{3})\Big|_{\theta=\pi}&=I_{\alpha\{\beta_{1}\beta_{2}\beta_{3}\}}(p_{1},p_{2},|p_{3}|)\nn\\&\quad
-i \pi p_{1}^{\beta_{1}}p_{2}^{\beta_{2}}p_{3}^{\beta_{3}}\int_{0}^{\infty}\D x\,x^{\alpha}K_{\beta_{1}}(p_{1}x)K_{\beta_{2}}(p_{2}x)I_{\beta_{3}}(|p_{3}|x),
\label{tripleKcont}
\end{align}
where the phase $e^{-i\pi\beta_3}$ from the continuation of the Bessel function cancels with that from the continuation of $p_3^{\beta_3}$. 
To evaluate the flat space limit of this analytically continued  integral, we now consider the asymptotic behaviour of its integrand.  
Physically, the flat space limit is reached by going to  $
p_ix \gg 1$, which corresponds to the deep interior of the bulk spacetime.
Replacing the Bessel functions with their asymptotic behaviours,
\[
K_{\beta}(p_i x)\rightarrow\sqrt{\frac{\pi}{2p_i x}}\,e^{-p_ix},\qquad I_{\beta}(p_i x)\rightarrow\sqrt{\frac{1}{2\pi p_i x}}\,e^{p_i x},
\]
we obtain the flat space limit
\begin{equation}
\lim_{E\rightarrow 0} I_{\alpha\{\beta_{1}\beta_{2}\beta_{3}\}}(p_{1},p_{2},p_{3})\Big|_{\theta=\pi}=
-\frac{\pi^{3/2}\Gamma(\alpha-1/2)\Pi_{i=1}^{3}p_{i}^{\beta_{i}-1/2}}{\sqrt{8}E^{\alpha-1/2}}.
\label{flat}
\end{equation}
Note this result derives entirely from the $KKI$ integral in \eqref{tripleKcont}, since the triple-$K$ integral is finite for collinear configurations \cite{Bzowski:2013sza}.
This asymptotic formula agrees with our result \eqref{flat1000} for 
the flat space limit of $I_{1\{000\}}$, as well as the results  in \cite{Farrow:2018yni}  for odd spacetime dimensions.

In even dimensions, however, we encounter divergent triple-$K$ integrals and hence we must take into account the effects of regularisation and renormalisation  
\cite{Bzowski:2015pba, Bzowski:2017poo, Bzowski:2018fql}. 
In general, a triple-$K$ integral diverges whenever the indices satisfy
\[\label{singcond}
\alpha+1 \pm \beta_1\pm \beta_2 \pm \beta_3 = -2n
\]
for any (independent) choice of $\pm$ signs and non-negative integer $n$.
To regulate, one performs infinitesimal shifts of the operator and spacetime dimensions, and thus of the indices $\alpha, \{\beta_i\}$ parametrising the triple-$K$ integrals.  The divergences can then be extracted and eliminated by the addition of 
covariant local counterterms, before removing the regulator to obtain the renormalised correlators.  

For the correlators of interest here, one finds from the detailed analysis of  \cite{Bzowski:2017poo, Bzowski:2018fql} that the regulated form factors contain only {\it ultralocal} divergences, meaning they are analytic functions of the squared momenta.\footnote{Such singularities correspond to triple-$K$ integrals with sign choice $(---)$ in \eqref{singcond}. In fact, for $\<TTT\>$, $\<JJ\O\>$ and $\<TT\O\>$, one also encounters individual triple-$K$ integrals with semilocal $(--+)$ or $(+--)$ divergences, however these either cancel with one another, or else are multiplied by vanishing  coefficients.  The regulated form factors then contain only ultralocal $(---)$ divergences.  This is consistent with the absence of $(--+)$ or $(+--)$ type counterterms for these correlators.}
Terms of this form, and the corresponding counterterm contributions, cannot contribute any singular behaviour in the flat space limit: this would require the appearance of factors of $E$ raised to negative powers, which are not ultralocal.
It therefore suffices to 
apply the continuation \eqref{tripleKcont} to the regulated form factors and extract the leading behaviour as $E\rightarrow 0$ using the asymptotic formula \eqref{flat}.  The result is necessarily finite as the regulator is  removed and all dimensions are restored to their physical values.\footnote{From the analysis of \cite{Bzowski:2015pba}, the $KKI$ integral in \eqref{tripleKcont} is singular only when the sign of $\beta_3$ in \eqref{singcond} is $+$.  However, as above, all singularities of this type  either cancel or are multiplied by vanishing coefficients.}

\subsection{\texorpdfstring{Analytic continuation of the renormalised form factors in $d=4$}{Analytic continuation of the renormalised form factors in d=4}}\label{zmethod}

Where the renormalised form factors are known explicitly, we can alternatively  apply the analytic continuation \eqref{continuation} directly to the renormalised form factors. 
 In this approach we never encounter any divergences since we always work within the renormalised theory.

To illustrate this we consider the case $d=4$, where
all renormalised  form factors have been evaluated in terms of differential operators acting on the finite master integral $I_{1\{000\}}$ \cite{Bzowski:2017poo, Bzowski:2018fql}.  
These derivatives can be evaluated by making repeated use of the relation
\begin{align}\label{Ireduce}
p_1\frac{\partial}{\partial p_1}I_{1\{000\}} &=\frac{1}{J^2}\Big[ 2p_1^2(p_1^2-p_2^2-p_3^2)I_{1\{000\}} \nn\\&\quad -
p_1^2\ln p_1^2+\frac{1}{2}(p_1^2+p_2^2-p_3^2)\ln p_2^2 +\frac{1}{2}(p_1^2-p_2^2+p_3^2)\ln p_3^2\Big].
\end{align}
Ultimately, one finds that all renormalised form factors are given by a linear combination of the master integral $I_{1\{000\}}$ multiplied by some rational function of the squared momenta and $\sqrt{-J^2}$, plus logarithms of the momenta and RG scale multiplied by similar rational functions, plus polynomials in the squared momenta.
As we now discuss, 
the analytic continuation of all these terms is easily accomplished using the results of section \ref{master}.  

The situation is clearest in the $(z,\zb)$ variables, where all factors of $\sqrt{-J^2}$ rationalise according to \eqref{Jsqz}. 
The renormalised form factors $A_n$ then take the form
\begin{align}\label{Azform}
p_3^{\ell}A_n =a_n^{(0)}p_3^2  I_{1\{000\}}+a_n^{(1)} \ln (z\zb)+a_n^{(2)} \ln\big( (1-z)(1-\zb)\big)+a_n^{(3)}\ln \frac{p_3^2}{\mu^2} + a_n^{(4)},
\end{align}
where the factor of $p_3^\ell$ renders $A_n$ dimensionless, and the $a_n^{(m)}$ are specific rational functions of $z$ and $\zb$.
In particular, these rational functions may diverge as $\zb\rightarrow z$.
Although we acquire new terms from analytically continuing the logs, the leading behaviour in the flat space limit in fact always comes from the continuation of the master integral $I_{1\{000\}}$.  
This can be understood as follows.
The key point is that the renormalised form factors are finite in the collinear limit,\footnote{In the regulated theory, the form factors are linear combinations of triple-$K$ integrals, and triple-$K$ integrals do not have collinear singularities \cite{Bzowski:2013sza}.  The counterterm contributions are ultralocal and thus do not have collinear singularities either.  The renormalised form factors are then finite for collinear configurations, as can be checked explicitly using the results of  \cite{Bzowski:2017poo, Bzowski:2018fql}.} which corresponds to sending $\zb\rightarrow z$ while keeping $\theta=0$.  
Since in the collinear limit \cite{Chavez:2012kn}
\[\label{I1000coll}
\lim_{\zb\rightarrow z} I_{1\{000\}}\Big|_{\theta=0} = -\frac{1}{2p_3^2}\Big(\frac{\ln z}{1-z}+\frac{\ln(1-z)}{z}\Big),
\]
we see that if $a_n^{(0)}$ diverges as $(z-\zb)^{-k}$ for some $k$, then $a_n^{(1)}$ and $a_n^{(2)}$ must also diverge at this same order, so that 
\[
\chi_1 = \lim_{\zb\rightarrow z} \Big(2a_n^{(1)}-\frac{1}{2(1-z)}a_n^{(0)}\Big),
\qquad
\chi_2 = \lim_{\zb\rightarrow z} \Big(2a_n^{(2)}-\frac{1}{2z}a_n^{(0)}\Big)
\]
are both finite.  The remaining rational functions $a_n^{(3)}$  and $a_n^{(4)}$ are both subleading: $a_n^{(3)}$ is finite  in the collinear limit, since  no cancellations are possible for this term, while any collinear divergences in $a_n^{(4)}$ are of order $(z-\zb)^{-k+2}$,
since they must cancel against the polynomial terms that arise in $I_{1\{000\}}$ at subleading  order 
({\it i.e.,} at order $(z-\zb)^2$ relative to the leading term shown in \eqref{I1000coll}).
 
After we analytically continue to $\theta=\pi$, the coefficient $a_n^{(0)}$ now acquires an additional factor of $(z-\zb)^{-1}$ from analytically continuing the master integral according to \eqref{zflatlimit}.
The continuations of the log terms in \eqref{Azform} do not produce any additional divergences, however, and so overall the leading  $(z-\zb)^{-k+1}$  behaviour of the form factor is that associated with the master integral. 
Thus, to find the leading behaviour of the renormalised form factors  in the flat space limit, we actually only need to know the coefficient $a_n^{(0)}$ of the master integral.    
This is a substantial simplification.

Finally,  let us  remark  that our approach 
is also readily applicable to the problem of continuing 
Euclidean CFT correlators to Lorentzian signature, as studied  in  \cite{Bautista:2019qxj,Gillioz:2019lgs,Anand:2019lkt}.  One simply needs to analyse how $z$ and $\zb$ move in the complex plane under Wick rotation, then evaluate the corresponding continuation of the master integral as discussed in section \ref{master}.

\subsection{Extracting the dependence on the master integral in general dimensions}
\label{reltoI1000}

For general even dimensions above four, results are available for the regulated form factors, and the nature of all divergences and counterterms have been tabulated \cite{Bzowski:2017poo, Bzowski:2018fql}.  
Once again, 
all  triple-$K$ integrals can be computed starting from the master integral $I_{1\{000\}}$ using the reduction scheme of \cite{Bzowski:2015yxv}.   
To obtain the final renormalised form factors then requires a certain amount of additional case-by-case analysis of counterterm contributions.  While this analysis is easy to perform in any specific case,  it is difficult to write down general closed-form expressions. 
From the $d=4$ discussion above, however, all we really need to know is the contribution to the renormalised form factors coming from the master integral $I_{1\{000\}}$, since this is the term that dominates in the flat space limit.  This contribution is easily evaluated as we now explain.

Working in the regulated theory to avoid divergences, all triple-$K$ integrals are first reduced to the integral $I_{0\{111\}}$.   This can be achieved  
using the relations \cite{Bzowski:2015yxv}
\begin{align}
I_{\alpha+1\{\beta_{1}+1,\beta_{2},\beta_{3}\}}&=\left(2\beta_{1}-p_{1}\partial_{p_{1}}\right)I_{\alpha\{\beta_{1}\beta_{2}\beta_{3}\}},
\label{op1}\\[1ex]
I_{\alpha+2\{\beta_{1},\beta_{2},\beta_{3}\}}&=\Big(\partial_{p_{j}}^{2}+\frac{1-2\beta_{j}}{p_{j}}\partial_{p_{j}}\Big)I_{\alpha\{\beta_{1}\beta_{2}\beta_{3}\}},
\label{op2}\\
I_{\alpha+1\{\beta_{1}+1,\beta_{2}+1,\beta_{3}+1\}}&=\frac{1}{\alpha-\beta_{t}-1}B_{\beta_1,\beta_2,\beta_3} I_{\alpha\{\beta_{1}\beta_{2}\beta_{3}\}}
\label{op3}
\end{align} 
where in \eqref{op2} one can choose any $p_j$ from $j=1,2,3$ and the operator in \eqref{op3} is
\[\label{BigBdef}
B_{\beta_1,\beta_2,\beta_3} = 
p_{1}^{2}\big(2\beta_{2}-p_{2}\partial_{p_{2}}\big)\left(2\beta_{3}-p_{3}\partial_{p_{3}}\right)+\mathrm{cyclic}.
\]
From these relations, we can construct the five index-shifting operations listed in Table \ref{operations}. The first three operations in this table  follow from cyclic permutations of  \eqref{op1}, while the fourth and fifth operation are  \eqref{op2} and \eqref{op3} respectively. 

From (4.2)\,-\,(4.7) of \cite{Bzowski:2015yxv}, 
the form of the dimensionally regulated integral $I_{0\{111\}}$ is 
\[
I_{0+u\ep\{1+v_1\ep,1+v_2\ep,1+v_3\ep\}} = \frac{I^{(-2)}}{\ep^2}+\frac{I^{(-1)}}{\ep} +I^{\mathrm{(scheme)}} + I^{\mathrm{(scale-violating)}}+I^{\mathrm{(nonlocal)}}+O(\ep),
\]
where the indices have been shifted by an infinitesimal parameter $\ep$ times scheme-dependent constants $(u,v_1,v_2,v_3)$.  The divergent terms $I^{(-2)}$ and $I^{(-1)}$ are ultralocal and semilocal respectively, and will ultimately be removed by subtracting counterterm contributions.  The scheme-dependent term $I^{\mathrm{(scheme)}}$ contains logarithms of the individual momentum magnitudes and hence is semilocal, while the scale-violating piece $I^{\mathrm{(scale-violating)}}$ contains products of such logarithms and is nonlocal.  The final scale-invariant, 
nonlocal piece $I^{\mathrm{(nonlocal)}}$ (also referred to as $I_{0\{111\}}^{(fin)}$ in \cite{Bzowski:2017poo, Bzowski:2018fql}) encodes the dependence on the master integral we seek:
\[
I^{\mathrm{(nonlocal)}}=
\frac{J^{2}}{4}I_{1\{000\}}.
\]
As in our discussion for $d=4$, the finiteness of the renormalised form factors in the collinear limit 
ensures the leading contribution in the flat space limit 
comes solely from $I^{\mathrm{(nonlocal)}}$.  Unlike $I_{1\{000\}}$, neither $I^{\mathrm{(scheme)}}$ or $I^{\mathrm{(scale-violating)}}$ acquire any additional divergences as $\zb\rightarrow z$ after continuing to $\theta=\pi$.  Since for $\theta=0$ any divergences as $\zb\rightarrow z$ must cancel, the leading divergence as $\zb\rightarrow z$ for $\theta=\pi$ must then come from $I_{1\{000\}}$.
For this reason, we can simply replace
\[
I_{0\{111\}} \rightarrow \frac{J^{2}}{4}I_{1\{000\}}
\label{0111to1000}
\]
for the purposes of computing the flat space limit of the renormalised form factors. 
All details of the regularisation scheme and renormalisation analysis can be safely neglected, since their contribution is subleading in the flat space limit.

\begin{table}
\begin{center}
\begin{tabular}{|c|c|c|c|c|}
\hline
operation & $\delta \alpha$ & $\delta \beta_1$ & $\delta \beta_1$ & $\delta \beta_3$ \\ \hline
{\bf 1} & 1 & 1 & 0& 0 \\ \hline
{\bf 2}  &1 & 0 &1& 0\\ \hline
{\bf 3}  &1 & 0 & 0&1 \\ \hline
{\bf 4}  & 2 & 0 & 0& 0\\ \hline
{\bf 5}  &1& 1 &1 &1\\ \hline
\end{tabular}
\end{center}
\caption{Index-shifting operations generated by cyclic permutations of \eqref{op1} (operations ${\bf 1}$, ${\bf 2}$ and ${\bf 3}$), along with \eqref{op2} and \eqref{op3} (operations ${\bf 4}$ and ${\bf 5}$).  Through repeated use of these operations, all the regulated triple-$K$ integrals  in even-dimensional correlators can be reduced to $I_{0\{111\}}$.}

\label{operations}
\end{table}

In summary, the known expressions for the regulated form factors can be related to $I_{0\{111\}}$ using the operations \eqref{op1}\,-\,\eqref{op3} summarised in Table 
\ref{operations}, after which we substitute \eqref{0111to1000}.  The leading behaviour in the flat space limit then corresponds to applying the same sequence of differential operators to the leading flat space behaviour of $(J^2/4)I_{1\{000\}}$, evaluated using  \eqref{flat1000}.
In fact, one only needs to keep track of the leading contributions when evaluating these derivatives, which enables further simplification. 
For example, the flat space limit of $I_{2\{111\}}$ can be obtained by applying operation {\bf{5}} in Table \ref{operations}, which simplifies to
\[
\lim_{E\rightarrow0}I_{2\{111\}}\propto\left(p_{1}^{2}p_{2}p_{3}+\mathrm{cyclic}\right)\partial_{E}^{2}\frac{1}{\sqrt{c_{123}E}}\propto\frac{\sqrt{c_{123}}}{E^{3/2}}.
\]
Here, we retained only the terms in \eqref{BigBdef} featuring derivatives, since their action is to generate more singular powers of $E$.  For this same reason, the factor of $\sqrt{c_{123}}$ from the flat space limit of $I_{1\{000\}}$ can be moved outside the derivatives, which can then be replaced by derivatives with respect to $E$ using the chain rule.
Using this method, we find that the triple-$K$ integrals we encounter all exhibit flat space behaviour in agreement with the asymptotic formula  \eqref{flat}.

\section{General even dimensions}

In this section, we apply the procedure of section \ref{reltoI1000} to compute the flat space limit of 3-point correlators of  stress tensors, currents and marginal scalar operators, which we denote $T$, $J$, and $\mathcal{O}$, respectively.  We show that correlators reduce to flat space scattering amplitudes in one higher dimension related by a double copy, extending the results of \cite{Farrow:2018yni} to even dimensions.
We take the spacetime dimension to be $d=2n>4$,  postponing the analysis of $d=4$ (where more detailed results for the renormalised form factors are available) to the following section.

\subsection{$\left\langle JJJ\right\rangle $}

First, we  consider the 3-point correlator of conserved currents. This can be decomposed into form factors as follows:
\begin{equation}
\left\langle J J J \right\rangle =A_{1}(p_{1},p_{2},p_{3})\,{\epsilon}_{1}\cdot {p}_{2}\,{\epsilon}_{2}\cdot {p}_{3}\, \epsilon_{3}\cdot p_{1}+\big[A_{2}(p_{1},p_{2},p_{3})\,\epsilon_{1}\cdot\epsilon_{2}\,\epsilon_{3}\cdot p_{1}+\mathrm{cyclic}\big].
\label{jjj}
\end{equation}
Here, $\<JJJ\>$ represents the correlator fully contracted with polarisation vectors, and with colour factors suppressed.  We also strip off the overall delta function associated with momentum conservation.\footnote{Such correlators are denoted $\lla \ldots \rra$ in \cite{Bzowski:2017poo, Bzowski:2018fql}.}
From the conformal Ward identities, one finds \cite{Bzowski:2017poo}
\begin{align}
A_1 = C_{1}I_{n+2\{n-1,n-1,n-1\}}, \qquad 
A_2  = C_{1}I_{n+1\{n-1,n-1,n\}}+C_{2}I_{n\{n-1,n-1,n-1\}},
\end{align}
where $C_1$ and $C_2$ are constants, and 
\[\label{C2JJJ}
C_2 = \# C_1 + \# C_{JJ},
\]
where $C_{JJ}$ is the normalisation of the 2-point function.
The $\#$ represent specific dimension-dependent constants whose precise form is not be important for reasons explained below.

\begin{figure}
\centering
	       \includegraphics[scale=.39]{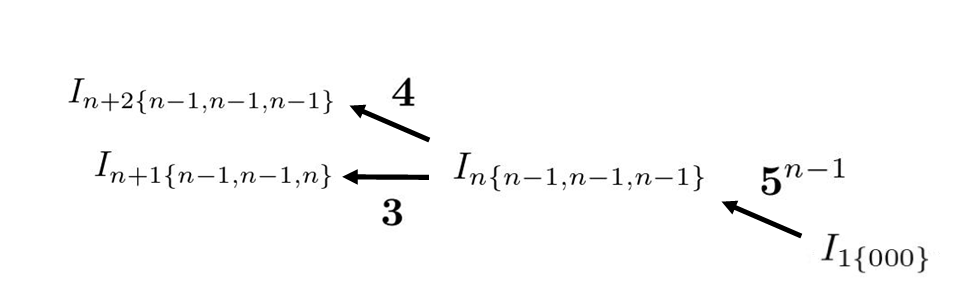}
    \caption{Reduction scheme for the regulated triple-$K$ integrals appearing in $\<JJJ\>$, where the numbered operations refer to  Table \ref{operations}, and ${\bf 5}^{n-1}$ means applying operation ${\bf 5}$ a total of $n-1$ times.} 
    \label{jjjr}
\end{figure}

The reduction to $I_{0\{111\}}$ and then to $I_{1\{000\}}$ via \eqref{0111to1000} is depicted in Figure \ref{jjjr}, where circled numbers correspond to operations in Table \ref{operations}.
Applying these operations to \eqref{flat1000} and using \eqref{C2JJJ},
 we can read off the flat space limit,
\begin{equation}
\ensuremath{\lim_{E\rightarrow0}\left\langle JJJ\right\rangle }\propto c_{123}^{(d-3)/2}\left[\frac{C_{1}}{E^{(d+3)/2}}\left(\mathcal{A}_{F^{3}}+\mathcal{O}(E)\right)+\frac{C_{JJ}}{E^{(d-1)/2}}\left(\mathcal{A}_{YM}+\mathcal{O}(E)\right)\right],
\label{flatjjj}
\end{equation}
where the gauge theory scattering amplitudes $\mathcal{A}_{F^{3}}$ and $\mathcal{A}_{YM}$ are given in \eqref{gaugeamp}.
In this calculation, the contribution from the first term in \eqref{C2JJJ} is subleading, while the $\#$ in the second term proportional to $C_{JJ}$ has been absorbed into the overall constant of proportionality in \eqref{flatjjj}.  Since $C_1$ is an arbitrary constant, we have additionally rescaled   $C_1$ to eliminate any relative factors between the two terms in \eqref{flatjjj}.

\subsection{$\left\langle TTT\right\rangle $}

The form factor decomposition for the stress tensor 3-point function is 
\begin{align}
\left\langle T T T \right\rangle  
&\nonumber =A_{1}(p_1,p_2,p_3)\left(\epsilon_{1}\cdot p_{2}\,\epsilon_{2}\cdot p_{3}\,\epsilon_{3}\cdot p_{1}\right)^{2}  \\ 
\nonumber &\quad +\big(A_{2}(p_1,p_2,p_3)\,\epsilon_{1}\cdot\epsilon_{2}\,\epsilon_{1}\cdot p_{2}\,\epsilon_{2}\cdot p_{3}\left(\epsilon_{3}\cdot p_{1}\right)^{2} + \mathrm{cyclic}\big)\\
\nonumber &\quad +\big(A_{3}(p_1,p_2,p_3)\left(\epsilon_{1}\cdot\epsilon_{2}\right)^{2}\left(p_{1}\cdot\epsilon_{3}\right)^{2}+\mathrm{cyclic}\big)\\
\nonumber &\quad +\big(A_{4}(p_1,p_2,p_3)\,\epsilon_{1}\cdot\epsilon_{3}\,\epsilon_{2}\cdot\epsilon_{3}\,\epsilon_{1}\cdot p_{2}\,\epsilon_{2}\cdot p_{3}+\mathrm{cyclic}\big)\\
&\quad +A_{5}(p_1,p_2,p_3)\,\epsilon_{1}\cdot\epsilon_{2}\,\epsilon_{2}\cdot\epsilon_{3}\,\epsilon_{3}\cdot\epsilon_{1},
\label{tttgeneral} 
\end{align}
where $\<TTT\>$ represents the correlator fully contracted with polarisation vectors and with the delta function of momentum conservation stripped off.
The form factors are  \cite{Bzowski:2017poo}
\begin{align}\label{TTTformfactorsgend}
A_1 & = C_{1}I_{5+n\{n,n,n\}}, \\
A_2 & = 4C_{1}I_{4+n\{n,n,n+1\}}+C_{2}I_{3+n\{n,n,n\}}, \\
A_3 & = 2C_{1}I_{3+n\{n,n,n+2\}}+C_{2}I_{2+n\{n,n,n+1\}}+C_{3}I_{n+1\{n,n,n\}},\\
A_4 & = 8C_{1}I_{3+n\{n+1,n+1,n\}}-2C_{2}I_{2+n\{n,n,n+1\}}+C_{4}I_{n+1\{n,n,n\}}, \\
A_5 & = 8C_{1}I_{n+2\{n+1,n+1,n+1\}}
+2C_{2}\big(I_{n+1\{n+1,n+1,n\}}\nn\\[0.5ex]& \qquad
+I_{n+1\{n+1,n,n+1\}}
+I_{n+1\{n,n+1,n+1\}}\big)
+C_{5}I_{n-1\{n,n,n\}},
\end{align}
where $C_1$ to $C_5$ are constants (independent of those introduced in the previous subsection) which are related by
\[ 
C_{4}=2C_{3}+\#C_{2},\quad C_{5}=\#C_{1}+\#C_{2}+\#C_{3}.
\]
Here, the $\#$ are specific dimension-dependent constants whose form is not of interest since the corresponding terms are subleading in the flat space limit.  The factor of two in the first equation is however important in order to recover the Einstein gravity amplitude.
We can likewise  replace $C_3$ in terms of the normalisation $C_{TT}$ of the stress tensor 2-point function,
\[
C_3 = \# C_1 +\# C_2 + \# C_{TT},
\]
where the terms proportional to $C_1$ and $C_2$ are also  subleading in the flat space limit.  

\begin{figure}
\centering
	       \includegraphics[scale=.36]{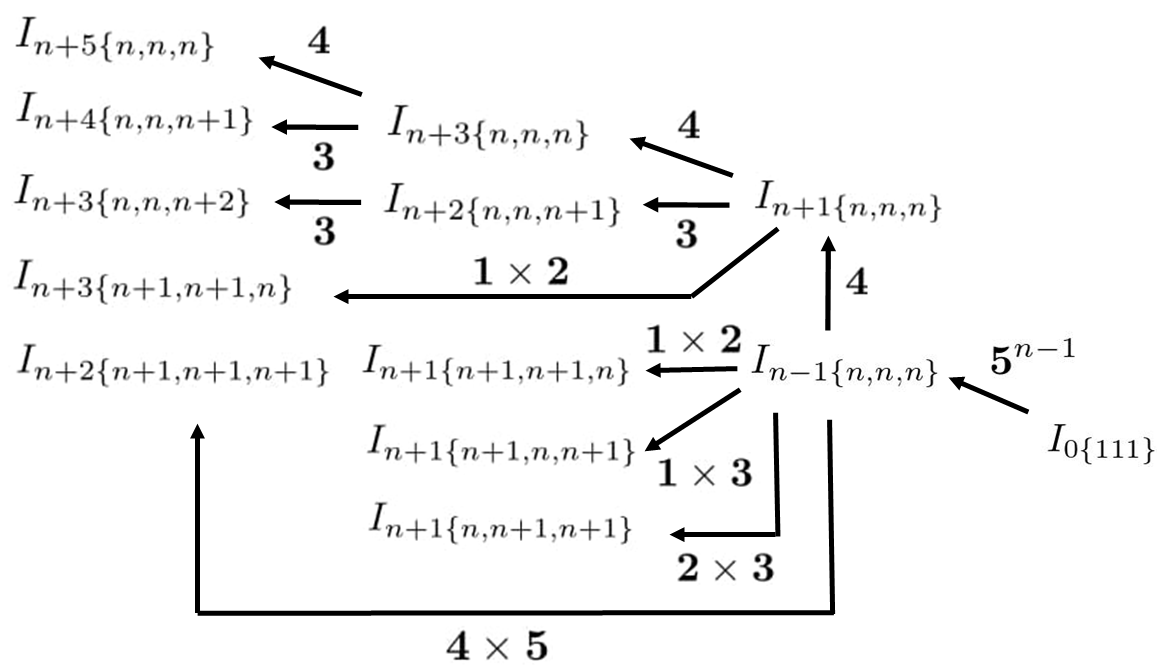}
    \caption{Reduction scheme for the regulated triple-$K$ integrals appearing in $\<TTT\>$, where the numbered operations refer to those in Table \ref{operations}.} 
    \label{tttr}
\end{figure}

The reduction of the triple-$K$ integrals to $I_{0\{111\}}$ is depicted in Figure \ref{tttr}. 
Using \eqref{0111to1000}, the flat space limit of these integrals can then be deduced from that of $I_{1\{000\}}$.  After re-scaling the (theory-specific) constants $C_1$ and $C_2$ to absorb dimension-dependent constants, we obtain the flat space limit
\begin{align}\label{flatttt}
\lim_{E\rightarrow0}\ensuremath{\left\langle TTT\right\rangle } &\propto c_{123}^{(d-1)/2}\left[\frac{C_{1}}{E^{(d+9)/2}}\left(\mathcal{A}_{W^{3}}+\mathcal{O}(E)\right)\right.\nn\\&\qquad\qquad\quad\quad
\left.+\frac{C_{2}}{E^{(d+5)/2}}\left(\mathcal{A}_{\phi R^{2}}^{222}+\mathcal{O}(E)\right)+\frac{C_{TT}}{E^{(d+1)/2}}\left(\mathcal{A}_{EG}+\mathcal{O}(E)\right)\right].
\end{align}
Remarkably, the gravitational amplitudes arising in the flat space limit of $\<TTT\>$ are double copies of the gauge theory amplitudes arising in the flat space limit of $\<JJJ\>$, as given in \eqref{gravityamp}.  
This result  takes the same form as in odd dimensions \cite{Farrow:2018yni}, as one would expect from the dimension-independent nature of the amplitudes themselves.

\subsection{$\left\langle JJO\right\rangle $}
For the case of two currents and a marginal scalar, the form factor decomposition  is 
\begin{equation}
\left\langle JJ\O\right\rangle =-A_{1}(p_{1},p_{2},p_{3})\,{\epsilon}_{1}\cdot{p}_{2}\,{\epsilon}_{2}\cdot{p}_{1}+A_{2}(p_{1},p_{2},p_{3})\,\epsilon_{1}\cdot\epsilon_{2}.
\label{jjogdcor}
\end{equation}
The form factors are  \cite{Bzowski:2018fql}
\begin{align}
A_{1}=C_{1}I_{n+1\{n-1,n-1,n\}}, \qquad 
A_{2}=C_{1}I_{n\{n-1,n-1,n+1\}}+C_{2}I_{n-1\{n-1,n-1,n\}},
\end{align}
where the constants $C_1$ and $C_2$ (which are once again independent from those defined in previous subsections) satisfy
\[
C_{2}=\# C_{1}.
\]

\begin{figure}
\centering
	       \includegraphics[scale=.36]{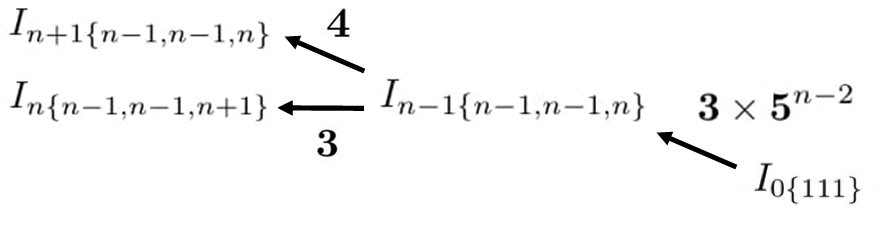}
    \caption{Reduction scheme for the regulated triple-$K$ integrals appearing in $\<JJ\O\>$.}
    \label{jjor}
\end{figure}

Using the reduction to $I_{1\{000\}}$ in Figure \ref{jjor},
we then obtain
\[\label{flatjjo}
\lim_{E\rightarrow0}
\left\langle JJ\O\right\rangle \propto c_{123}^{(d-3)/2}p_3 \frac{C_1 }{E^{(d+1)/2}} \Big(\mathcal{A}_{\phi F^2}+\mathcal{O}(E)\Big),
\]
where $\mathcal{A}_{\phi F^2}$ is the amplitude for two gluons and a scalar in \eqref{dilaton}.

\subsection{$\left\langle TTO\right\rangle $}
Finally, in the case of two stress tensors and a marginal scalar, we have
\begin{align}
\left\langle TT\O\right\rangle &=A_{1}(p_1,p_2,p_3)\,\left(\epsilon_{1}\cdot p_{2}\,\epsilon_{2}\cdot p_{1}\right)^{2}\nn\\&\quad  -A_{2}(p_1,p_2,p_3)\,\epsilon_{1}\cdot\epsilon_{2}\,\epsilon_{1}\cdot p_{2}\,\epsilon_{2}\cdot p_{1}+A_{3}(p_1,p_2,p_3)\left(\epsilon_{1}\cdot\epsilon_{2}\right)^{2},
\label{ttogdcor}
\end{align}
where the form factors are \cite{Bzowski:2018fql}
\begin{align}
A_1 & = C_{1}I_{3+n\{n,n,n\}}, \\
A_2 & = 4C_{1}I_{2+n\{n,n,n+1\}}+C_{3}I_{n+1\{n,n,n\}}, \\
A_3 & = 2C_{1}I_{n+1\{n,n,n+2\}}+C_{2}I_{n\{n,n,n+1\}}+C_{3}I_{n-1\{n,n,n\}},
\end{align}
and the constants satisfy
\[
C_{2}=\#C_{1},\qquad C_{3}=\#C_1.
\]

\begin{figure}
\centering
	       \includegraphics[scale=.4]{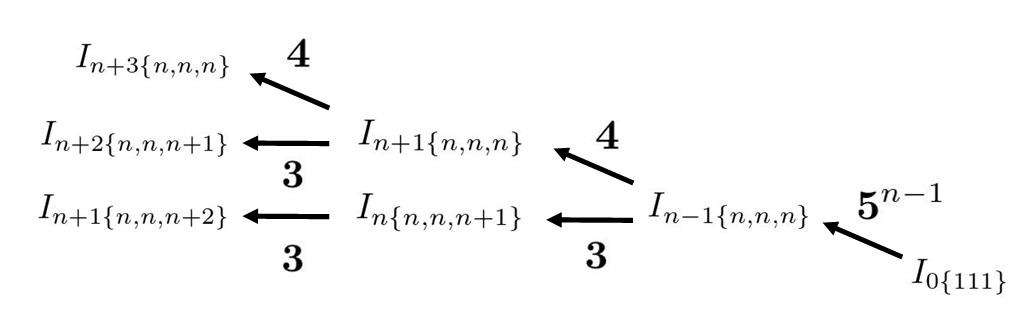}
    \caption{Reduction scheme for the regulated triple-$K$ integrals appearing in $\<TT\O\>$.  All numbered operations refer to those in  Table \ref{operations}.} 
    \label{ttor}
\end{figure}

Using the triple-$K$ reduction in Figure \ref{ttor},
the flat space limit is given by
\[\label{flattto}
\lim_{E\rightarrow0}\left\langle TT\O\right\rangle \propto c_{123}^{(d-1)/2} \frac{C_{1}}{E^{(d+5)/2}}\left(\mathcal{A}_{\phi R^{2}}^{220}+\mathcal{O}\left(E\right)\right)
\]
where the amplitude $\mathcal{A}_{\phi R^{2}}^{220}$ for two gravitons and a scalar is a double copy of that for two gluons and a scalar arising in the flat space limit of $\<JJ\O\>$, as given in  \eqref{ymdc}.  Once again, we find the same double copy structure for even-dimensional correlators as that obtained in \cite{Farrow:2018yni} for odd-dimensional correlators.

\section{Four dimensions} \label{4d}

We present the case of $d=4$ separately since the complete renormalisation analysis has been carried out
in \cite{Bzowski:2017poo, Bzowski:2018fql}.
For $\<TTT\>$ this allows us to parametrise the flat space limit in terms of the trace anomaly coefficients. We also briefly discuss the double copy structure for anomalies.

\subsection{$\left\langle JJJ\right\rangle $}

In $d=4$, the renormalised form factors for $\left\langle JJJ\right\rangle $ are given by \cite{Bzowski:2017poo}
\begin{align}
A_1 & = - C_1 p_1 p_2 p_3 \frac{\partial^3}{\partial p_1 \partial p_2 \partial p_3} I_{1\{000\}}+\ldots, \\
A_2 & =C_1 p_1 p_2 p_3^2 \frac{\partial^2}{\partial p_1 \partial p_2} I_{1\{000\}} + 4C_{JJ} I_{2\{111\}}^{\text{(fin)}} +\ldots,
\end{align}
where the ellipses denote terms which are non-singular in the flat space limit.  We have additionally suppressed a factor relating to the  colour and the charge multiplying the 2-point normalisation $C_{JJ}$. 
The finite integral $I_{2\{111\}}^{\mathrm{(fin)}}$ is given in (3.48) of \cite{Bzowski:2017poo}.
Using \eqref{flat1000} and \eqref{op1} - \eqref{0111to1000}, we recover the flat space limit in \eqref{flatjjj}.  

\subsection{$\left\langle TTT\right\rangle $}

The renormalised form factors for $\left\langle TTT\right\rangle $ are \cite{Bzowski:2017poo}
\begin{align}\label{A1TTTren}
A_1 & = C_1 I_{7\{222\}}+\ldots, \\
A_2 & =  2 \Big[a + c - 2 C_1 p_3 \frac{\partial}{\partial p_3} \Big] I^{\text{(fin)}}_{5\{222\}}+\ldots,\\
A_3 & = 2 \Big[ 2 c - (a + c + C_1) p_3 \frac{\partial}{\partial p_3} + C_1 p_3^2 \frac{\partial^2}{\partial p_3^2} \Big] I_{3\{222\}}^{\text{(fin)}}+\ldots,\\
A_4 & = 4 \Big[ c-a + (a+c) p_3 \frac{\partial}{\partial p_3}+2 C_1 \Big( 8 - 4 \sum_{j=1}^3 p_j \frac{\partial}{\partial p_j} + p_1 p_2 \frac{\partial^2}{\partial p_1 \partial p_2} \Big) \Big] I_{3\{222\}}^{\text{(fin)}}+\ldots,\\
A_5 & = 2(a+c) \Big[ 32 - 8 \sum_{j=1}^3 p_j \frac{\partial}{\partial p_j} + 2 \sum_{i<j} p_i p_j \frac{\partial^2}{\partial p_i \partial p_j} \Big] I_{1\{222\}}^{\text{(fin)}} \nn\\&\quad 
- 8 C_1 p_1^3 p_2^3 p_3^3 \frac{\partial^3}{\partial p_1 \partial p_2 \partial p_3} I_{1\{000\}} 
 +\ldots,
\end{align}
where once again the ellipses denote terms that are non-singular in the flat space limit.  Here, the triple-$K$ integral $I_{7\{222\}}$ and the finite integrals $ I_{5\{222\}}^{\text{(fin)}}$, $I_{3\{222\}}^{\text{(fin)}}$ and $I_{1\{222\}}^{\text{(fin)}}$ can all be expressed as derivatives of the master integral, see (3.198)\,-\,(3.201) of \cite{Bzowski:2017poo}.  The coefficients 
$a$ and $c$ are those entering the trace anomaly,
\[\label{tranomaly}
\left\langle T_{\mu}^{\mu}\right\rangle =a E_{4}+cW^{2}, 
\] 
where $E_4$ is the Euler density and $W^2$ is the square of the four-dimensional Weyl tensor.  
(Note that the Euler coefficient $a$ is often defined with an additional minus sign to here.) 
To ensure the 2-point function is traceless, we work in a scheme where the $\Box R$ contribution to the trace anomaly vanishes.

Using \eqref{flat1000} and \eqref{op1} - \eqref{0111to1000}, we recover the flat space limit in \eqref{flatttt} with\footnote{The exact relation from \cite{Bzowski:2017poo} is $C_{TT} = -2c$, but here as 
in \eqref{flatttt} we omit such numerical coefficients.} 
\[\label{replacements}
C_2\rightarrow a+c, \qquad C_{TT}\rightarrow  c.
\] 
We can equivalently write this as
\begin{align}\label{4dTTTflat1}
&\lim_{E\rightarrow0}\frac{E^{13/2}}{c_{123}^{3/2}}\left\langle TTT\right\rangle \propto C_1 \mathcal{A}_{W^{3}}, \\[1ex]\label{4dTTTflat2}
&\lim_{E\rightarrow0}\frac{E^{9/2}}{c_{123}^{3/2}}\left.\left\langle TTT\right\rangle \right|_{C_{1}=0}\propto (a+c) \mathcal{A}_{\phi R^{2}}^{222}, \\[1ex]\label{4dTTTflat3}
&\lim_{E\rightarrow0}\frac{E^{5/2}}{c_{123}^{3/2}}\left.\left\langle TTT\right\rangle \right|_{C_{1}=0, \, a+c=0}\propto c\, \mathcal{A}_{EG}.
\end{align}
The coefficients of the $\phi R^2$ and Einstein gravity amplitudes arising in the flat space limit  are thus parametrised by the trace anomaly.  This  is natural from a holographic perspective since these anomaly coefficients are determined by bulk gravitational interactions  
\cite{Henningson:1998gx,Nojiri:1999mh,Bugini:2016nvn}. 
Hence for Einstein gravity, for example, 
$a+c = 0$ and the  $\mathcal{A}_{\phi R^2}^{222}$ contribution vanishes.

Finally, while our focus 
is on the transverse traceless parts of correlators, 
we note that double copy structure 
 also arises in 
 the trace part of this correlator
  as shown in  \cite{Bzowski:2017poo}. 
From the trace Ward identity, this takes the form
\begin{align}
\<T^\mu_\mu(\bs{p}_1)T(\bs{p}_2)T(\bs{p}_3)\> &=2\<T(\bs{p}_2)T(-\bs{p}_2)\>+2\<T(\bs{p}_3)T(-\bs{p}_3)\>+\mathcal{A},
\end{align}
where we trace over the first two indices and contract the rest with polarisation tensors.
The anomalous contribution $\mathcal{A}$ comes from functionally differentiating the trace anomaly \eqref{tranomaly}.
The part proportional to the Euler anomaly coefficient $a$
 is then a double copy of the chiral anomaly:
\[\label{Eulerdoublecopy}
\mathcal{A}_{Euler}=40\, a \,\mathcal{A}_{chiral}^2, \qquad
\mathcal{A}_{chiral}=\epsilon_{\mu_{2}\mu_{3}\mu_{4}\mu_{5}}\epsilon_{2}^{\mu_{2}}\epsilon_{3}^{\mu_{3}}p_{2}^{\mu_{4}}p_{3}^{\mu_{5}},
\]
where the chiral anomaly arises in the transverse Ward identity for  
currents,
\begin{align}
&\<
\left(p_{1}\cdot J^{a}(\bs{p}_1)\right)J^{b}(\bs{p}_2)J^{c}(\bs{p}_3)\big\rangle \nn\\[1ex]
&\qquad \qquad = g f^{adc}\<J^{d}(\bs{p}_2)J^b(-\bs{p}_2)\>- g f^{abd}\<J^{d}(\bs{p}_3)J^c(-\bs{p}_3)\>+d^{abc}\mathcal{A}_{chiral}.
\end{align}
In this identity, $f^{abc}$ is the structure constant, $g$ the gauge coupling, and $d^{abc}$ is a group-theoretic factor depending on the matter content.  
The double copy \eqref{Eulerdoublecopy} 
derives from the specific structure of type A anomalies (in the classification of \cite{Deser:1993yx}), and 
is not present for type B anomalies such as the Weyl-squared contribution to the trace anomaly.

\subsection{$\left\langle JJO\right\rangle $}
The renormalised form factors are given by \cite{Bzowski:2018fql}
\begin{align}
A_{1}  =C_{1}\left(2-p_{3}\frac{\partial}{\partial p_{3}}\right)I_{2\{111\}}^{\text{(fin)}}+\ldots, \qquad 
A_{2}  =C_{1}p_{3}^{2}I_{2\{111\}}^{\text{(fin)}}+\ldots,
\end{align}
where the omitted terms are non-singular in the flat space limit. 
Using \eqref{flat1000} and \eqref{op1}\,-\,\eqref{0111to1000}, we recover the flat space limit in \eqref{flatjjo}. 

\subsection{$\left\langle TTO \right\rangle $}
The renormalised form factors are given by \cite{Bzowski:2018fql}
\begin{align}
A_{1}  &=C_{1}\left(2-p_{1}\frac{\partial}{\partial p_{1}}\right)\left(2-p_{2}\frac{\partial}{\partial p_{2}}\right)\left(2-p_{3}\frac{\partial}{\partial p_{3}}\right)I_{2\{111\}}^{\text{(fin)}}+\ldots,\\
A_{2}  &=4C_{1}\left(1-p_{3}\frac{\partial}{\partial p_{3}}\right)I_{3\{222\}}^{\text{(fin)}}+\ldots\\
A_3 & = 2C_{1}I_{3\{222\}}^{\text{(fin)}}+\ldots,
\end{align}
where again the omitted terms are non-singular in the flat space limit.
Using \eqref{flat1000} and \eqref{op1}\,-\,\eqref{0111to1000}, we recover the flat space limit in \eqref{flattto}. 

\section{Conclusion} \label{concl}

In this paper, we extended to even spacetime dimensions our results  for the double copy structure of momentum-space CFT 
correlators \cite{Farrow:2018yni}.  
This double copy  structure is inherited from the bulk scattering amplitudes that arise on taking the flat space limit of correlators.  
Our main achievement is to understand the analytic continuation required to reach the flat space limit. 
Analytically continuing the largest momentum magnitude, which is an energy from the bulk perspective, we arrive at
configurations with $E=0$ for which bulk energy conservation is restored.
Analysing the behaviour of the master integral $I_{1\{000\}}$ under this continuation, the flat space limits of all renormalised correlators can then  be constructed. 
 
Prior to analytic  continuation, the master integral can be expressed as a 1-loop triangle integral and evaluated in terms of a Bloch-Wigner function in suitable complex variables. Under analytic continuation, these complex variables follow a simple path in the complex plane which involves crossing two branch cuts. The resulting discontinuities produce a new term which supplies the necessary singular behaviour  in the flat space limit. This term can also be derived from  the leading singularity of the 1-loop triangle integral, revealing interesting connections to dual conformal symmetry. The flat space limit of all other triple-$K$ integrals, and that of the correlators themselves, can then be deduced by applying differential operators to $I_{1\{000\}}$.

The above discussion holds for 3-point correlators of stress tensors, currents, and marginal scalars of general CFTs in all even dimensions greater than two, and we verified it explicitly it in four dimensions where the renormalised correlators have been fully evaluated. 
In this case, 
we showed that anomalies play an important role in the flat space limit. In particular, we found that the flat space limit of stress tensor correlators is controlled by conformal anomalies, in 
line with general holographic expectations.

It is remarkable that double copy structure plays such a ubiquitous role in correlation functions of general CFTs. It would be interesting to explore how this extends to higher-point correlators in momentum space. A general solution to the conformal Ward identities for $n$-point scalar correlators in momentum space was recently proposed in \cite{Bzowski:2019kwd}, so it would be interesting to extend this to tensorial  correlators and understand how to systematically compute their flat space limit. Note that the general solution in \cite{Bzowski:2019kwd} can be written as a 3-loop Feynman integral so it is conceivable that the flat space limit is encoded in the leading singularity of this integral. It may also be fruitful to look for double copy structure in the correlators derived from Witten diagrams of specific theories in the bulk such as bi-adjoint scalars, Yang-Mills,  and Einstein gravity. KLT-like relations for inflationary graviton correlators have been explored in \cite{Li:2018wkt}, and  our results for double copy structure can be 
likewise  applied to cosmology. Finally, it would  be of interest to explore if the analytic continuation we used to reach the flat space limit can be adapted to continue CFT correlators from Euclidean to Lorentzian signature.  This problem has recently been analysed via other methods in  \cite{Bautista:2019qxj,Gillioz:2019lgs,Anand:2019lkt}, but our approach here seems particularly promising.

\bigskip

\subsection*{ Acknowledgements}

AL is supported by a Royal Society University Research Fellowship. PM is supported by an Ernest Rutherford Fellowship from the Science \& Technologies Facilities Council.

\appendix

\section{Leading singularity of the master integral} \label{Leadingsing}
\label{leadingsing_app}

The master integral $I_{1\left\{ 000\right\} }$ for four-dimensional 3-point CFT correlators is a limit of the dual conformal box integral \cite{Broadhurst:1993ib, Usyukina:1992jd}.  This box integral also plays a prominent role in the context of $\mathcal{N}=4$ SYM 
\cite{Brandhuber:2008pf,Drummond:2008vq}. The leading singularity can easily be computed by writing the box integral in coordinates which make the dual conformal symmetry manifest (the region momentum coordinates), then evaluating the global residue.

We begin by writing the box integral as
\[
\Phi(u,v) = x_{13}^2 x_{24}^2
\int\D^4 {\bf x}_5\, \frac{1}{x_{15}^2 x_{25}^2 x_{35}^2 x_{45}^2}, 
\]
where $\x_{ij}=\x_i-\x_j$ and the region momentum coordinates are related to the external momenta by
\[
\x_{12}=\bs{p}_{1},\qquad \x_{23}=\bs{p}_{2},\qquad \x_{34}=\bs{p}'_{3},\qquad \x_{41}=\bs{p}'_{4}.
\]
The integral is invariant under translations and inversions $\bs{x}_i \rightarrow \bs{x}_i/x_i^2$, and therefore has conformal symmetry in region momentum space, known as dual conformal symmetry \cite{Drummond:2006rz}. As a result, it depends only on the dual conformal cross-ratios 
\[
u =\frac{x_{12}^2 x_{34}^2}{x_{13}^2 x_{24}^2}, \qquad
v = \frac{x_{14}^2 x_{23}^2}{x_{13}^2 x_{24}^2}.
\]

\begin{figure}
\centering
	       \includegraphics[scale=0.9]{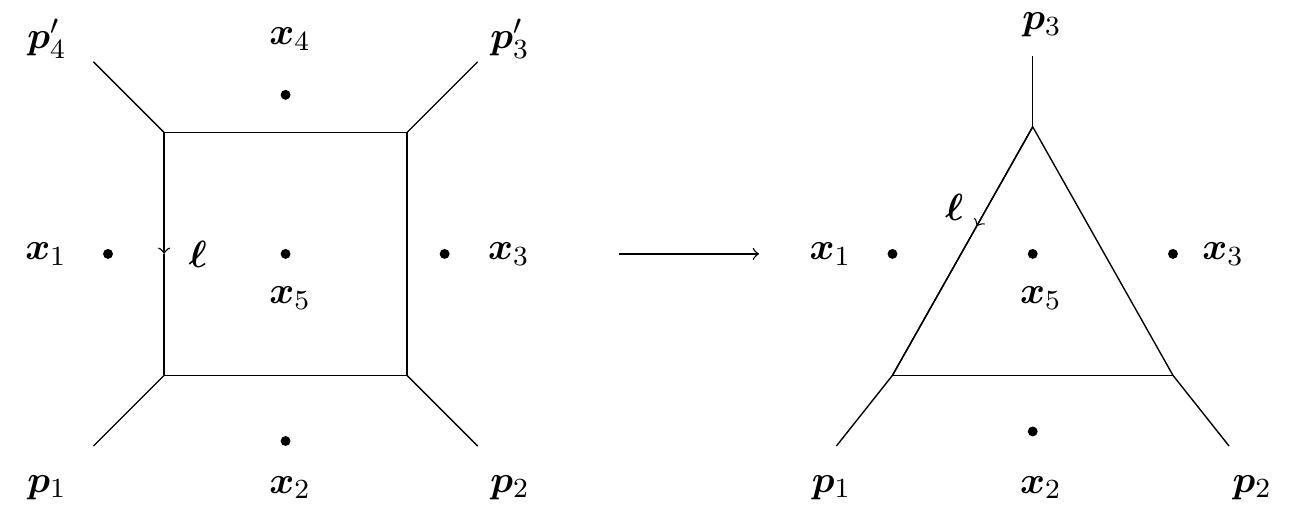}
    \caption{Region momenta relating the massless box to the 3-point master integral $I_{1\{000\}}$.  All external momenta are taken as ingoing.} 
    \label{loops}
\end{figure}

To recover a 3-point function in momentum space, we  define
\[
\bs{x}_{12} = \bs{p}_1, \qquad \bs{x}_{23}=\bs{p}_2, \qquad \bs{x}_{31} = \bs{p}_3, 
\]
where $\bs{p}_3 = \bs{p}'_3 + \bs{p}'_4$, and take the limit $\x_4\rightarrow \infty$.  We then recover 
\[
u = \frac{p_1^2}{p_3^2}, \qquad v = \frac{p_2^2}{p_3^2},
\] 
and setting $\bs{l}=\bs{x}_{51}$, we find
\begin{align}
\lim_{x_{4}\rightarrow\infty}\Phi(u,v)&=
\int\D^{4}{\bf x}_{5}\,\frac{x_{13}^{2}}{x_{15}^{2}x_{25}^{2}x_{35}^{2}}=\int\D^{4}\bs{\ell}\,\frac{p_{3}^{2}}{\ell^{2}(\ell+p_{1})^{2}(\ell-p_{3})^{2}} =4\pi^{2}p_{3}^{2}I_{1\{000\}}.
\end{align}
The box and triangle integrals in region momentum space are depicted in Figure \ref{loops}.

An efficient method to evaluate the leading singularity of the box integral was devised in \cite{Drummond:2013nda}. In place of $x_5^\mu$, we change variables to the four new coordinates 
\[
P_i = x_{i5}^2, \qquad  i=1,\ldots,4.  
\]
The Jacobian for this transformation is 
\[
\mathcal{J} = \det\Big(\frac{\partial P_i}{\partial x_5^\mu}\Big) = \det (-2x_{i5}^\mu).
\]
As $\det{(\bs{M} \bs{M}^T)}=(\det \bs{M})^2$, taking $M_i^\mu = -2 x_{i5}^\mu$ we have
\[
\mathcal{J}^2  = \det(4 x_{i5}\cdot x_{j5}) = 2^4 \det(x_{ij}^2-x_{i5}^2-x_{j5}^2),
\]
since $\bs{x}_{ij} = \bs{x}_{i5}-\bs{x}_{j5}$ and the matrix is $4\times 4$.
We now have
\[
\int\D^4 {\bf x}_5\, \frac{1}{x_{15}^2 x_{25}^2 x_{35}^2 x_{45}^2} = \int \frac{\D^4P_i}{\mathcal{J}}\frac{1}{P_1 P_2 P_3 P_4},
\]
and the leading singularity is just the global residue
\[
(2\pi i)^{4}\frac{1}{\mathcal{J}}\Big|_{P_{i}=0}=4\pi^{4}\frac{1}{\sqrt{\det{x_{ij}^{2}}}}=\frac{4\pi^{4}}{x_{13}^{2}x_{24}^{2}(z-\bar{z})}.
\]
The leading singularity of $I_{1\left\{ 000\right\} }$ is then
\[
\frac{\pi^2}{ p_3^2 (z-\bar{z})} = \frac{\pi^2}{\sqrt{-J^2}},
\]
which agrees with \eqref{flat1000}.

\bibliographystyle{JHEP}

\bibliography{dsamp}

\end{document}